\documentclass[amssymb,prl,aps,twocolumn,amsmath,showpacs]{revtex4}

\usepackage[dvips]{graphicx}
\usepackage{amstext}

\begin{document}

\title{Observation of spin-triplet superconductivity in Co-based Josephson Junctions}
\author{Trupti S. Khaire, Mazin A. Khasawneh, W. P. Pratt, Jr., Norman O. Birge}
\email{birge@pa.msu.edu}
\affiliation{Department of Physics and
Astronomy, Michigan State University, East Lansing, Michigan
48824-2320, USA}
\date{\today}

\begin{abstract}

We have measured a long-range supercurrent in Josephson junctions
containing Co (a strong ferromagnetic material) when we insert
thin layers of either PdNi or CuNi weakly-ferromagnetic alloys
between the Co and the two superconducting Nb electrodes. The
critical current in such junctions hardly decays for Co
thicknesses in the range of 12-28 nm, whereas it decays very
steeply in similar junctions without the alloy layers.  The
long-range supercurrent is controllable by the thickness of the
alloy layer, reaching a maximum for a thickness of a few nm. These
experimental observations provide strong evidence for induced
spin-triplet pair correlations, which have been predicted to occur
in superconducting/ferromagnetic hybrid systems in the presence of
certain types of magnetic inhomogeneity.

\end{abstract}

\pacs{74.50.+r, 74.45.+c, 75.70.Cn, 74.20.Rp} \maketitle

When a conventional spin-singlet superconductor is brought into
contact with a normal metal, superconducting pair correlations
penetrate into the normal metal over distances as large as a
micron at low temperature, creating the superconducting proximity
effect \cite{Deutscher}.  If the normal metal is replaced by a
ferromagnet, the pair correlations penetrate only a few
nanometers, as the exchange field in the ferromagnet leads to a
rapid loss of phase coherence between electrons with
opposite-pointing spins \cite{Buzdin-JETP, Demler}.  This
limitation would not arise if the Cooper pairs in the
superconductor had spin-triplet symmetry, which occurs only rarely
in nature \cite{Mackenzie, Saxena}.  It was predicted several
years ago that spin-triplet superconducting correlations could be
induced at the interface between a conventional spin-singlet
superconductor and a ferromagnet with inhomogeneous magnetization
\cite{Bergeret:01a,Kadigrobov:01}.  Moreover, these pair
correlations are in a new symmetry class: they have even relative
orbital angular momentum but are odd in frequency or time
\cite{Bergeret:05}. A promising hint of spin-triplet correlations
in half-metallic CrO$_{2}$ was reported in 2006 by Keizer
\textit{et al.} \cite{Keizer}; however there has been no
confirmation of that result in the intervening time.  Here we
present strong evidence for spin-triplet pair correlations in
Josephson junctions fabricated from common metals: Nb and Co.  The
magnetic inhomogeneity is supplied by thin layers of a
weakly-ferromagnetic alloy -- either PdNi or CuNi -- inserted
between the Co and Nb layers.  As the Co thickness is increased,
the maximum supercurrent in the Josephson junctions decays very
slowly -- in sharp contrast to the very fast decay observed in
similar junctions without these alloy layers \cite{Khasawneh}. The
strength of the triplet correlations can be controlled by the
thickness of the alloy layer, reaching its maximum for a thickness
of a few nm.

A schematic diagram of our Josephson junction samples is shown in
Figure 1a.  The entire multilayer structure up through the top Au
layer is sputtered onto a Si substrate in a single run, without
breaking vacuum between subsequent layers. The multilayers are
subsequently patterned into circular pillars using
photolithography and Ar ion milling, after which the SiO${_x}$
insulating layer is thermally evaporated to isolate the top Nb
contact from the base. Finally the top Nb contact is sputtered
through a mechanical mask. The Au layer is fully superconducting
due to the proximity effect with the surrounding Nb layers.  The
Nb superconducting layers have critical temperature near 9 K,
which allows us to measure the Josephson critical supercurrent at
4.2 K with the samples dipped in liquid helium.  Details of our
fabrication and measurement procedures are given in our previous
publications \cite{Khasawneh,Khaire}.

\begin{figure}[tbh]
\begin{center}
\includegraphics[width=3.0in]{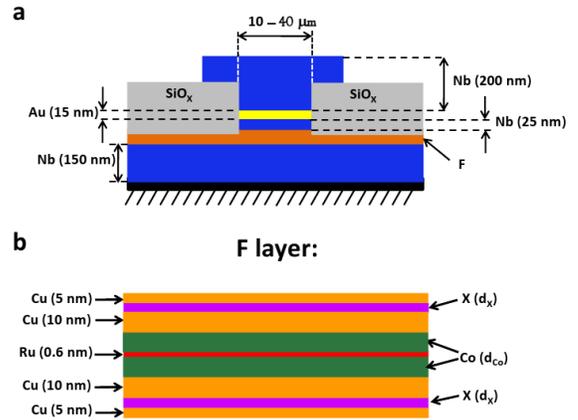}
\end{center}
\caption{(color online).  \textbf{a}) Schematic diagram of the
Josephson junction samples, shown in cross-section.  \textbf{b})
Detailed sequence of the metal layers inside the Josephson
junctions (labelled F in a). The layers labelled X are either PdNi
or CuNi alloy. The functions of the various layers are described
in the text. Only the thicknesses of the Co and X layers are
varied in this work. The Cu buffer layers play no active role in
the devices, but are important to isolate the X layers
magnetically from the Co layers.}\label{Schematic}
\end{figure}

The detailed sequence of internal layers (labelled F for
``ferromagnetic" in Figure 1a) is shown in Figure 1b.  The layers
labelled X represent either Pd$_{0.88}$Ni$_{0.12}$ or
Cu$_{0.48}$Ni$_{0.52}$ ferromagnetic alloys.  The purpose of the
Co in the central Co/Ru/Co trilayer is to suppress the
conventional spin-singlet Josephson supercurrent.  The critical
current of similar Josephson junctions containing Co/Ru/Co
trilayers, but without the X layers, decays exponentially with
increasing Co thickness with a decay constant of $2.34 \pm 0.08$
nm \cite{Khasawneh}.  The thin Ru layer induces antiparallel
exchange coupling between the domains in the two Co layers
\cite{Parkin:90}, to produce nearly zero total magnetic flux in
the junctions.  As a result, the critical current vs. applied
magnetic field data of these Josephson junctions exhibit
nearly-ideal ``Fraunhofer patterns," as shown in Figure 2.  These
patterns give us reliable measurements of the maximum possible
critical current in each sample, while also indicating that the
current flow in the junctions is uniform and that there are no
shorts in the surrounding SiO insulator.

\begin{figure}[tbh]
\begin{center}
\includegraphics[width=2.8in]{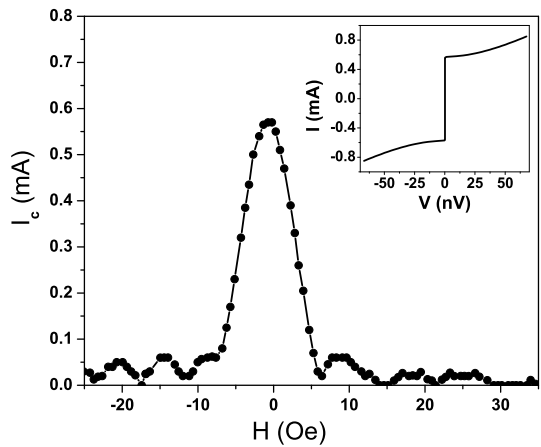}
\end{center}
\caption{Critical current ($I_c$) vs. applied magnetic field ($H$)
for a 10 $\mu$m diameter Josephson junction with $d_{Co}$ = 13 nm
and $d_{PdNi}$ = 4 nm, measured at $T = 4.2$ K. The excellent
``Fraunhofer pattern" results from cancellation of the intrinsic
magnetic flux in the junction, due to antiparallel exchange
coupling of the two Co layers via the thin Ru layer.  (The lines
are guides to the eye.)  The inset shows the current-voltage
($I-V$) characteristic of the junction at $H=0$.}\label{IV-Fraun}
\end{figure}

We discuss first the case where X = Pd$_{0.88}$Ni$_{0.12}$, a
weakly-ferromagnetic alloy with a Curie temperature of 175 K
\cite{Khaire}.  Figure 3a shows the product of critical current
and normal state resistance, $I_c R_N$, vs. total cobalt
thickness, $D_{Co}\equiv$ 2$d_{Co}$, for a series of samples with
fixed PdNi layer thickness, $d_{PdNi} = 4$ nm.  (The normal state
resistance, $R_N$, is determined from the inverse slope of the
$I-V$ curve for $I \gg I_c$.) There is no discernible decay of
$I_c R_N$ for $D_{Co} > 12$ nm.  For comparison, Figure 3a also
shows data from Ref. \cite{Khasawneh} for junctions not containing
PdNi. In those samples $I_c R_N$ decays very rapidly with
increasing $D_{Co}$.  When $D_{Co} = 20$ nm, $I_c R_N$ is over 100
times larger in the samples with PdNi than in the samples without
PdNi. The long-range character of the Josephson current in samples
with PdNi represents strong evidence for its spin-triplet nature.

\begin{figure}[tbh]
\begin{center}
\includegraphics[width=3.2in]{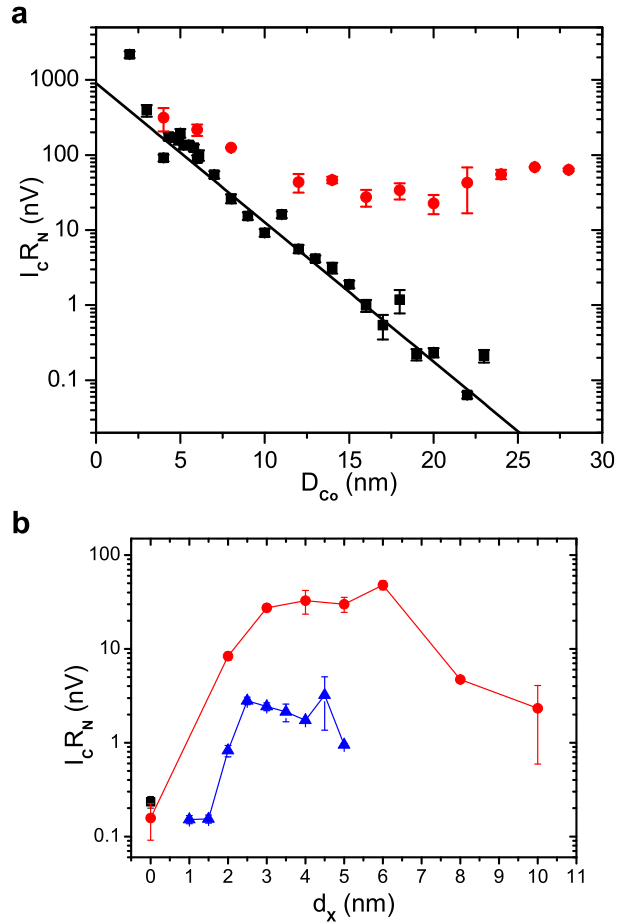}
\end{center}
\caption{(color online). \textbf{a}) Product of critical current
times normal state resistance, $I_c R_N$, as a function of total
Co thickness, $D_{Co}$ = 2$d_{Co}$.  Red circles represent
junctions with X = PdNi and d$_{PdNi}$ = 4 nm, whereas black
squares represent junctions with no X layer (taken from Ref.
\cite{Khasawneh}).  As $D_{Co}$ increases above 12 nm, $I_c R_N$
hardly drops in samples with PdNi, but drops very rapidly in
samples without.  (The solid line is a fit of the data without
PdNi to a decaying exponential, also from Ref. \cite{Khasawneh}.)
Error bars represent the standard deviation determined from
measurements of two or more pillars on the same substrate, or are
set to 10$\%$ in the few cases where data from only a single
pillar were available. \textbf{b,}) $I_c R_N$ product as a
function of $d_X$ for two series of junctions with fixed $D_{Co}$
= 20 nm. Red circles: X = PdNi; blue triangles: X = CuNi. (The
square at $d_{X}$ = 0 is taken from Ref. \cite{Khasawneh}.) In
both cases, $I_c R_N$ first increases, then eventually decreases
with increasing $d_{X}$. Lines are guides to the eye.}\label{Main}
\end{figure}

The subtle role of the X layers in enhancing the supercurrent is
illustrated in Figure 3b, which shows $I_c R_N$ vs. $d_{X}$ with X
= PdNi or CuNi for two sets of samples with $D_{Co}$ fixed at 20
nm.  Without any X layer, $I_c R_N$ is very small, consistent with
the data shown in Figure 3a.  When the X layer reaches a critical
thickness, $I_c R_N$ increases rapidly, indicating that
spin-triplet pair correlations are being produced in the Nb/Cu/X
interface region.  $I_c R_N$ reaches a maximum for $d_{X}$ values
of a few nm, then decreases at larger values of $d_X$.  We believe
that the decrease in $I_c R_N$ at large $d_{X}$, visible for X =
PdNi, is caused by the destruction of the spin-triplet
correlations created at the Nb/Cu/X interface due to spin memory
loss in the bulk of the X layers.  The spin memory lengths in
these PdNi and CuNi alloys are very short -- about 2.8 nm in PdNi
\cite{Arham} and 1.4 nm in CuNi \cite{Oboznov:06}. This would
explain why we found no evidence for spin-triplet supercurrent in
our previous measurements of Josephson junctions containing only
PdNi layers of thickness 30-100 nm \cite{Khaire}. Evidently, a
thin PdNi or CuNi layer is essential to produce spin-triplet
correlations, whereas a thick layer suppresses them.

What essential properties give PdNi and CuNi their ability to
produce spin-triplet pair correlations in these Josephson
junctions? We speculate that the two crucial ingredients for
generation of the triplet are domain size, which should be
comparable to the superconducting coherence length in Nb, and
out-of-plane magnetocrystalline anisotropy.  While the domain size
of PdNi is not known, the domain size in Cu$_{0.47}$Ni$_{0.53}$
has recently been measured to be about 100 nm \cite{Veshchunov},
which is not so different from the Nb coherence length $\xi_S =
14$ nm. Competition between out-of-plane magnetocrystalline
anisotropy and the in-plane shape anisotropy of thin films can
lead to stripe domains with canted magnetization \cite{Koikeda}
and thus to non-collinear magnetizations in neighboring domains --
a key requirement for production of the triplet
\cite{Bergeret:05}.  Both PdNi \cite{Khaire} and CuNi
\cite{Ruotolo} are known to have out-of-plane magnetic anisotropy.
In this context, it is interesting to note what happens when the
Cu buffer layers between the X and Co layers are omitted.  We have
tried this for X = PdNi, and found that the supercurrent is much
smaller than in samples with Cu buffer layers.  Presumably the
domain structure of the PdNi is changed by exchange-coupling to
the Co, in a way that is detrimental to production of the triplet
correlations.  We suspect that the PdNi magnetization is forced to
lie in-plane near the interface with Co, which leads to less
non-collinear magnetization in neighboring domains \cite{Arham}.

It is natural to ask whether there are other materials besides
PdNi or CuNi that can produce spin-triplet correlations.  Clearly
Co alone is not sufficient, as demonstrated by the samples without
X layers \cite{Khasawneh}.  Scanning electron microscopy with
polarization analysis (SEMPA) measurements on Co films grown under
similar conditions as ours reveal magnetic domains with typical
sizes of a few microns, but with the magnetization directions of
neighboring domains largely antiparallel \cite{Borchers}.
Non-collinear magnetization resides only in the domain walls,
which is apparently not enough to produce a significant amount of
spin-triplet.  Aside from the presence of non-collinear
magnetizations, theory suggests that any "spin-active" interface
between a superconductor and a ferromagnet can produce
spin-triplet correlations \cite{Eschrig}.  We have tried using X =
Cu$_{0.94}$Pt$_{0.06}$, an alloy with strong spin-orbit
scattering, but preliminary data show very little, if any,
signature of the triplet.  We predict that PdCo, another weakly
ferromagnetic alloy with properties similar to those of PdNi, will
produce triplet correlations.

Comparison of our results with theory is problematic.  The
magnitude of the spin-triplet supercurrent depends on the details
of the PdNi or CuNi domain structure, while theoretical
calculations exist only for idealized magnetic configurations.
More useful is a discussion of the decay lengths of the
spin-singlet and spin-triplet supercurrents. In the ``dirty"
limit, where the mean free path, $l_e$, is the shortest relevant
length scale in the problem, the spin-singlet supercurrent should
decay on the length scale $\xi_F = \surd\overline{\hbar
D_F/E_{ex}}$, where $D_F$ and $E_{ex}$ are the electron diffusion
constant and exchange energy in the ferromagnet. Josephson
junctions containing Co, however, are in the ``intermediate"
limit, with $l_{e}$ longer than $\xi_F$, but shorter than $\xi_S$,
the superconducting coherence length. In that limit, the
spin-singlet supercurrent decays on the length scale $l_e$, which
is estimated to be 2.4 - 3.0 nm from previous studies
\cite{Khasawneh,Robinson:06}. Spin-triplet supercurrent, in
contrast, should decay over a much longer length scale given by
the smaller of the normal metal coherence length, $\xi_N =
\surd\overline{\hbar D_F/2\pi k_{B}T}$, or the spin memory length,
$L_{sf}=\surd\overline{\hbar D_F/\tau_{sf}}$, where $\tau_{sf}$ is
the mean time between spin-flip or spin-orbit scattering events.
Estimation of $D_F$ for Co is difficult due to its strong
ferromagnetism and to the widely-varying densities of states and
Fermi velocities of the different bands.  From our measured Co
resistivity, the Einstein relation, and the densities of states of
majority and minority electrons at the Fermi surface
\cite{Papaconstantopoulos}, we estimate $D_{F}=5\times10^{-3}$
m$^{2}$/s and $5\times10^{-4}$ m$^{2}$/s for the majority and
minority electrons, respectively, which give $\xi_N=40$ nm and 10
nm at $T=4.2$ K. $L_{sf}$ in Co has been measured to be about 60
nm, also with large uncertainty \cite{Piraux,Bass:07}.
Unfortunately, sample-to-sample fluctuations in the experimental
data in Figure 3a mask any discernible decay for $D_{Co}$ between
12 and 28 nm. Better statistics or data over a much larger range
of $D_{Co}$ will be needed to extract a meaningful estimate of the
decay length for the spin-triplet supercurrent.  We have not yet
attempted to increase $D_{Co}$ much more due to concerns about the
efficacy of the Co/Ru/Co exchange-coupled trilayer.

The spin-triplet pair correlations observed here and discussed in
Ref. \cite{Bergeret:05} are quite different from those believed to
occur in materials such as Sr$_2$RuO$_4$ \cite{Mackenzie}.  The
Cooper pairs in the latter satisfy the Spin-Statistics Theorem of
quantum mechanics by having odd relative orbital angular momentum
(p-wave). According to theory \cite{Bergeret:05}, the triplet pair
correlations induced in superconductor/ferromagnet hybrid systems
have even relative orbital angular momentum; in particular, they
can be s-wave, which implies that they are robust in the presence
of disorder. Quantum mechanics is not violated because the
correlations are odd in frequency, or equivalently odd under time
reversal.  This idea, first proposed in a model for liquid
helium-3 by Berezinskii \cite{Berezinskii}, is counter-intuitive,
as it implies that the equal-time pair correlation function
vanishes. Theoretical guidance is needed to find a direct method
to probe the odd-frequency aspect of the pair correlations
experimentally, rather than relying on the observed long-range
supercurrent to deduce the spin-triplet character combined with
robustness to disorder.

Looking back, there were hints of long-range proximity effects in
superconducting/ferromagnetic hybrid systems as early as 10 years
ago \cite{Giroud, Lawrence, Petrashov, Pena}, but there was no way
to control the observed effects.  More recently, Sosnin \textit{et
al.} \cite{Sosnin:06} observed phase-coherent oscillations in the
resistance of a Ho wire connected to two superconductors, but the
authors did not observe a Josephson supercurrent, nor did they
comment on its absence.  The observation by Keizer \textit{et al.}
\cite{Keizer} of a supercurrent through CrO$_2$ was an exciting
advance, but the critical currents in those samples varied by two
orders of magnitude in similar samples.  We anticipate that our
results, which exhibit systematic dependence of the spin-triplet
supercurrent on PdNi or CuNi thickness, will pave the way to many
new experiments \cite{Kontos:01, SanGiorgio}.

Acknowledgments:  We thank S. Bergeret, P. Brouwer, K.B. Evetov,
M. Stiles, and A. Volkov for helpful discussions, R. Loloee, B.
Bi, and Y. Wang for technical assistance, and use of the W.M. Keck
Microfabrication Facility.  This work was supported by the U.S.
Department of Energy under grant DE-FG02-06ER46341.

\end{document}